\documentstyle[11pt,iau185,twoside,epsf]{article}
\newcommand{\pol}{{1\over2}}
\begin{document}
\title{Changes in solar oscillation frequencies during the current activity
maximum:  analysis and interpretation}
\author{W. A. Dziembowski}
\affil{Warsaw University Observatory and Copernicus Center,
Poland}
\author{Philip R. Goode}
\affil{Big Bear Solar Observatory, New Jersey Institute of
Technology, U.S.A}

\begin{abstract}
We describe systematic changes in the  centroid frequencies and the
splitting coefficients as found using data from MDI on board SOHO,
covering cycle 23. The data allow us to
construct a seismic map of the evolving solar activity -- covering
all latitudes.  At lower latitudes, the temporal evolution closely
tracks that of {\it butterfly diagram}.  The additional information
from
higher latitudes in the map is of a significant activity in the polar
region, peaking at activity minimum in 1996.
The most
plausible source of solar oscillation frequency changes over the solar cycle is
the evolution of the radial component of the small-scale magnetic field.
The amplitude of the required mean field changes is $\sim 100$ G at the
photosphere, and increasing going inward.
\end{abstract}


\section{Phenomenology of frequency variations}

Using satellite
intensity data from cycle 21,  Woodard \& Noyes (1985)
first noted that p-mode frequencies increase with increasing solar
activity.  This discovery was confirmed by a number
of investigations made during cycle 22.
In particular, it was then established that the most significant
changes occur in the antisymmetric part of the spectrum of oscillation
multiplets -- that is, the part
that reflects the asphericity of the sun
(Kuhn, 1988; and Libbrecht \& Woodard, 1990). The dominant part of the
multiplet structure is
symmetric and arises from advection by rotation.

Oscillation data, from the on-going spatially resolved observations,
are represented as centroid frequencies, $\bar\nu_{\ell n}$ and
splitting coefficients, $a_{2k,\ell n}$, in the
following expression for the frequencies of the individual modes
\begin{equation}
\nu_{\ell nm}=\bar\nu_{\ell n}+\sum_{k=1} a_{k,\ell n} {\cal
P}_{k,\ell}.
\end{equation}
The quantities ${\cal P}$ are orthogonal
polynomials of $m$ defined for $2k\le\ell$ (see
Ritzwoller and Lavely 1991 and Schou, et al. 1994).

We use MDI data,
which contains coefficients up to $k=18$ for about 2000 p-
and f-modes with $\ell\le300$. There are 24 sets of data
corresponding to 72 day long measurements done between May 1996,
when the sun was at its activity minimum, to June 2001, when the
sun was in it high activity phase.

The relative changes of solar frequencies are of order
$10^{-4}$, which does is comparable to the individual measurement
errors. Significant rates of change are obtained by binning the
data or by assuming, as  Libbrecht and Woodard (1990) did, that the changes
scale inversely to mode inertia, which what is expected if the activity
related changes acts only near the surface. Following
this idea, we write the frequency changes in the form
\begin{equation}
 \bar\nu_{\ell,n}-(\bar\nu_{\ell,n})_{\rm
min}={\gamma_0\over I_{\ell,n}},
\end{equation}
where the subscript ``min" refers to measurements made at solar minimum.
We assume the same scaling for
the variable part of the $a$ coefficients of
even orders, and thus, we write
\begin{equation}
a_{2k,\ell,n}=a_{2k,\ell,n;{\rm rot}}+C_{k,\ell}{\gamma_k\over
I_{\ell,n}},
\end{equation}
The  numerical factor $C_{k,\ell}$ (see Dziembowski, et al, 1999, for the
explicit expression and its justification) was introduced to make
each $\gamma_k$ an unbiased probe of the
$P_{2k}$ distortion of the sun.
Here, we removed the constant contribution from the centrifugal force,
which is the only non-negligible effect of rotation.

The constant values of $\gamma_k$ inferred from the p-mode data
show systematic evolution as solar activity varies. The
coefficients, $\gamma_k$, up to $k=9$ are in excellent correlation with the
corresponding coefficients from the Legendre polynomial expansion of
the Ca II K line intensities, which are regarded as a good proxy
for the magnetic field (Dziembowski et al. 2000).

For p-modes, we expect that the $\gamma$'s do not depend on $\ell$,
if their source is localized near the surface. {\it A priori}, we
might expect a significant frequency dependence, but it
is not very strong.  This tells us something the precise localization of
the source.
For the f-modes, where we have the approximate
proportionality, $\nu\propto\sqrt{\ell}$, a  component of
$\Delta\nu$ was determined, which is also $\propto\sqrt{\ell}$ and grows with
increasing activity. This component may be
interpreted as evidence for a contraction of the sun's outer layers as activity rises.
Such contraction would
take place if the increase of the field were dominated by its
radial component (Dziembowski. Goode \& Schou, 2001)

\section{Seismic map of the sun's activity}

Having determined the $\gamma_k$ coefficients as functions of time,
we can construct a
seismic map of varying solar activity.  To do this, we determine the
quantity,
\begin{equation}
\gamma(\theta,t)=\sum_{k=0}\gamma_k(t)P_{2k}(\cos\theta).
\end{equation}
In Figure 1, we show the changes in $\gamma$ averaged into bins
$5\deg$-wide in latitude. At lower latitudes, we recognize
features that are well-known from the {\it butterfly diagram}: activity
appearing first at $35 - 45\deg$, and gradually moving toward the
equator.
However, there is also unexpectedly large activity in the polar
regions.  This decreases with the rise of activity at the lower
latitudes. This result is not an entirely surprising one, because it has been
known for some time that the polar field flips at
activity maximum. Furthermore, Moreno-Insertis \& Solanki
(1999) have already found that their models for low $\ell$
mode frequency behavior, during the activity cycle requires,
fields at high-latitude.
\begin{figure}
\begin{center}
\mbox{\epsfxsize=0.8\textwidth\epsfysize=0.53\textwidth\epsfbox{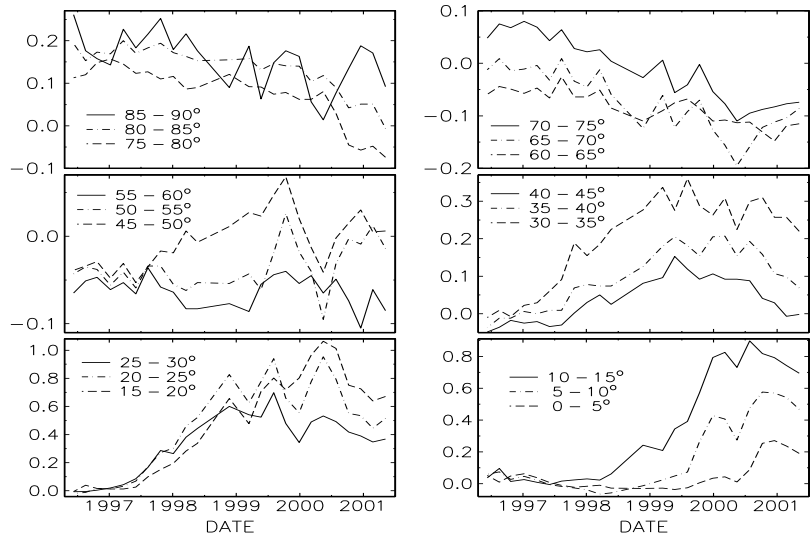}}
\caption{Zonal averaged values of $\gamma$ in $\mu$Hz. The ranges
of the latitude are given in the legend, and the overall
zero-point is arbitrary.}
\end{center}
\end{figure}

\section{The cause of the frequency changes}

There is ample evidence that the behavior of the $\gamma$s  in some way reflects
changes in the magnetic field in the outermost layers of the sun.
We would like is to infer more about the field: its
intensity and its behavior as a function of the depth below the
photosphere. For this it is essential to consider the role of
changes in the temperature and turbulent velocity induced by the
magnetic field.
A significant role of the latter effect may be eliminated (Dziembowski
and Goode, in preparation).

Following Goldreich et al. (1991), we consider a small-scale, random
magnetic field as the
primary cause of the frequency changes in which we allow the vertical
($r$) component to be statistically different from the two
horizontal ($\theta$ and $\phi$) ones.
In this, the averaged values $\overline{B_j^2}$ are treated as
functions of depth, and slowly varying functions of the co-latitude,
The latter dependence is represented in the form of a truncated
Legendre polynomial series,
\begin{equation}
\overline{B_iB_j}=\delta_{ij}\sum_{k=0}
[\delta_{jr}{\cal M}_{r,k}(r)+\pol{\cal M}_{H,k}(\delta_{j\theta}+
\delta_{j\phi}]P_{2k}(\cos\theta),
\end{equation}
where we included only seismically detectable (symmetric about equator) terms.

Each of the $k$-components gives rise to a $P_{2k}$ distortion
of the sun's structure.  For $k>0$, the hydrostatic equilibrium
condition suffices to determine the distortion of all the
thermodynamical parameters.

In the evaluation of the $\gamma$s, which in general must be treated
as functions of both $\nu$ and $\ell$,  we use
the variational principle for stellar
oscillations in which we treat the effects of the magnetic field as a
small perturbation. This principle, with use of hydrostatic
equilibrium and equations for adiabatic oscillations, leads to

\begin{equation}
\gamma_k=\int\left({\cal K}_{k,T}{\Delta T\over T}+{\cal K}_{k,r}^B
{\cal M}_{r,k}+{\cal K}_{k,H}^B{\cal M}_{H,k}\right)dD,
\end{equation}
where $D$ is depth. All the kernels, ${\cal K}$, may be explicitly expressed
in terms of parameters of the standard solar model and the radial
eigenfunctions of its p-modes.
If the magnetic perturbation is significant only in
the layers well-above the lower turning points
of all p-modes considered, then the kernels are both $k$ and $\ell$
independent.

Goldreich et al.(1991) considered only changes in centroid frequencies. They
pointed out that to explain
the frequency increase
between 1986 and 1988 a 1\% decrease of photospheric temperature is
needed (${\cal K}_{k,T}$ is
always $<0$). Regarding this requirement as being incompatible
with observations, they adopted the changing magnetic field as
the sole cause of the frequency increase; they found
that the field increase must be about 250 G at the photosphere,
and steadily growing to
about 1 kG at a depth of 10 Mm. Their numbers refer to the case of a
statistically isotropic field (${\cal M}_{H,k}=2{\cal M}_{r,k}$).
A much more modest field increase  ($<100$ G at the photosphere) would
result for an inwardly growing, pure radial field.
A similar result was obtained
by us (Dziembowski, Goode \& Schou, 2001) from our analysis of the MDI data.

For $k>0$, the term involving temperature change may be eliminated.
In the outer layers, the signs of the modified kernel are
${\cal K}_{k,r}^B>0$ and ${\cal K}_{k,H}^B<0$. Thus, the most economical
requirement is a pure radial field. With
the assumption that all the frequency changes are due to an
increase of the radial component of the small-scale magnetic fields,
the plots in Fig. 1 may be scaled from $\mu$Hz to Gauss using the
factor $\approx (190 {\rm G})^2/1 \mu{\rm Hz}$.

\begin{acknowledgments}
Research of WAD is supported by KBN grant 5P03D 012 20 and that of PRG by
NASA-NAG5-9682 and NSF-ATM-00-86999.
\end{acknowledgments}

\end{document}